\documentclass[11pt,twoside]{article}
\usepackage{newpasp}
\usepackage{epsf}

\index{Pfenniger, D.}

\begin{document}

\title{Maximum Galactic Disks vs. Hot Dark Halos}

\author{Daniel Pfenniger}
\affil{Geneva Observatory, University of Geneva,
CH-1290 Sauverny, Switzerland}

\begin{abstract}
A series of arguments is presented for heavy galaxy disks not only in
the optical regions, but also in the dark matter dominated regions of
spirals.  We are testing this possibility with extreme maximum disk
$N$-body models without any conventional spheroidal dark halo.
\end{abstract}

\keywords{galaxies, evolution of galaxies, dark matter}

\section{The Dark Matter in Galaxies and the Shape of Dark Halos}
The almost polar orbits of the Magellanic and Sagittarius streams, and
the existence of polar ring galaxies, suggest that the potential
quadrupolar term of the Milky Way and other spirals must be
significant.  But the determination of the corresponding density
flattening remains delicate, e.g., from a hot component tracing the
equipotentials such as X-ray gas, since potentials are usually much
rounder than the mass distribution (the potential of a thin disk with
flat rotation curve has still an axis ratio of about 0.7).

However other arguments suggest that the dark halo of spirals may
depart substantially from the {\it ellipsoidal\/} shape family almost
exclusively contemplated up to now. Namely, as we argue below, a
strong rotational support in the dark matter component offers a
solution for reconciling a series of problems raised recently in the
literature.

\subsection{Bright Spirals}
In the optical regions of bright spirals, several observational and
numerical works lead to an almost vanishing need of dark matter there.

The maximum disk hypothesis (e.g., van Albada et al. 1985) is
insufficient to prove that the optical disks are self-gravitating,
since the stellar $M/L$ ratio is insufficiently well constrained.  But
this hypothesis is consistent with observations.  On the other hand,
the wiggles in both the rotation curves and the luminous mass put some
constraints on the temperature of the supposed dark halo since a
massive hot component should damp the fluctuations seen in the
kinematically cold component.  However, in practice the projection
effect and non-circular shapes and velocities complicate the test.

Independently of this argument, several authors (e.g., Fux 1997; Gyuk
1999) have noted that the numerous Milky Way bulge micro-lensing
events can hardly be explained without an essentially maximum inner
disk of star like objects.  Other studies of Milky Way models
compatible with a large body of observations also tend to favor a
maximal disk (e.g., Sackett 1997).  It is interesting to note that
Dehnen \& Binney (1998) find that parametric mass models tend to
produce {\it hollow\/} halos, which they reject since they assume
implicitly hot, pressure supported halos.

The recent discovery of brown dwarfs in Orion (Zapatero Osorio et
al. 1999; Lucas \& Roche 2000) increases the likely mass estimate of
detected baryons in stars by perhaps 10--30\%, so is insufficient to
explain the dark matter in spirals. However, since brown dwarfs form
presumably in proportion to stars, the newly detected dark matter
component occurs precisely in the optical regions where little space
is left for another dark matter component.  Thus, the discovery of a
sizable fraction of ``brown stellar matter'' with a steeper profile
($\Sigma_\star \propto {\rm e}^{-R/h}$) than dark matter ($\Sigma_{\rm DM}
\propto R^{-1}$) increases the difficulty to fit a dark halo with a
density maximum in the region where it is in least demand.

Using constraints from dynamics, Debattista \& Sellwood (1998) or
Quillen \& Frogel (1997) find that bars are better compatible with
heavy optical disks and a small dark matter content.  A too massive
halo in the bar region brakes and destroys bars in a short time,
while the detailed features associated with the bar resonances would be
different with a fat dark halo that observed.

According to Dubinski et al. (1999), the length and shapes of tidal
tails produced in $N$-body models of galaxy interactions suggest that
the real tidal tails of interacting galaxies are mostly produced by
galaxies with disk-dominated rotation curves and low concentration
halos. This argument might perhaps be stronger with a wider
exploration of the possible galaxy shapes, in particular with extreme
dark matter disks as introduced here.

\subsection{Faint Galaxies}
In faint spiral galaxies the dark matter fraction is much higher than
in bright galaxies, but in contrast the ratio of HI/dark matter
varies little (Bosma 1981; Carignan \& Purton 1998).  Thus strangely
enough dark matter behaves differently that the stellar baryons but
knows very well about the gaseous ones.  Since gas eventually becomes
stars, and gas poor (or star rich) galaxies have little dark matter
too, it is natural to suggest that at least a substantial fraction of
it is a hard to detect phase of molecular hydrogen and helium (e.g.,
cold H$_2$ in gaseous, liquid, or solid forms), as we and others have
discussed a few years ago.

Since dominated by dark matter, faint spirals are suited to probe
the predictions resulting from CDM cosmological $N$-body simulations.
But the predicted steep central cusp of CDM halos appears incompatible with
the observed rotation curves (e.g., Moore et al. 1999).

Otherwise, the general presence of warps, spirals and asymmetries in
the outer gaseous disks of spirals is a strong indication that the
dark matter component there is not stabilizing much the disks.  In
particular the existence of a bar and large scale spiral arms in the
dark matter dominated gaseous disk of NGC 2915 (Bureau et al. 1999)
indicate to us a rather self-gravitating disk.

\subsection{Gas Infall}
The old hypothesis of secular gas infall in spirals has been recently
revived in the context of the also old observational problem raised by
the high-velocity clouds (HVCs) (Blitz et al. 1999).  These clouds
have also a high dark matter content in proportion to HI resembling
the one found in faint galaxies.  So if these clouds accrete onto the
Milky Way at a rate sufficient to produce the observed HI, then the
Milky Way dark matter could come simultaneously from these clouds (a
few $M_\odot/$yr of HI translates into a few tens $M_\odot/$yr of dark
matter which becomes comparable to the galaxy total mass after 10 Gyr). 

The important point to note about such a scenario is that the angular
momentum content of HI and dark matter should be also the same.  If
dark matter has a high angular momentum content then there is no
ground to suppose that the shape of the halos should be ellipsoidal.
On the contrary, all the known highly rotating gravitating structures
(e.g., protostellar disks, or rotating polytropes) adopt shapes 
like flaring disks or thick tori.

\subsection{Dark Halo Shapes}
If we remember the original motivations for introducing first
spherical, and later spheroidal dark halos, we see that over the years
most of them have disappeared.  The original motivation was to prevent
galaxies to make bars as easily as $N$-body simulations of disks did
(e.g., Ostriker et al. 1974).  The association of the dark halo with
the then thought round and virialized stellar halo was natural.  But
today bars are understood to be the rule in spirals, including the
Milky Way, while stellar halos appear rather non-virialized and made
of the leftover streams from dissolved dwarf spheroidals and globular
clusters.

The argument using Oort's constraint on the local mass density in the
Milky Way disk leads to require about twice as much density in a
fatter halo in view of the disk rotation velocity. But this constraint
depends largely on the assumption that the local density at the
Sun is representative of the azimuthally averaged density at the
Sun radius.  In the past, most of the Milky Way models have been built
on the assumption of strict axisymmetry, but today the Milky Way is
known to be barred, and the arm--inter-arm density contrast in other
spirals, as observed in the infrared, may be 2 or more!  Thus Oort's
constraint appears weaker today than before as long as the Sun position
with respect to the Milky Way major spiral arms and their effective 
amplitude are not better known.

In contrast to hot spheroidal halos, rotation supported dark halos
offer better prospectives to satisfy these constraints: 1) the inner
maximal optical disks, 2) the Bosma HI-DM relationship, 3) large
angular momentum accretion from the HVC's, and 4) flat rotation
curves.  Furthermore, unlike in ellipsoidal halos, in disk-like halos
the gravitational force from the outside on the inside is strong, so
disk-like halos are well ``connected'' by the mutual forces of each
parts, the ``halo-disk conspiracy'' is non-existent.  For a large
class of non-ellipsoidal halo potentials producing flat rotation
curves, see de Zeeuw \& Pfenniger (1988).

\section{N-body Models of Maximum Galactic Disks}
With the above motivations in mind, we have undertaken to test the
possibility of massive disks with $N$-body simulations.  Since the
rotation curves constrain little the 3D halo shapes except their
average mass density profile, it appears obvious that, say, half
rotation supported halos might easily satisfy the present constraints.
Instead, it is more challenging to push the rotation support to the
extreme limit, which is: 1) the scale-height of the dark component is
comparable to the one observed in the baryons (HI or stars), 2) the
{\it time average\/} pressure support corresponds to a marginally
stable Safronov-Toomre parameter $Q\approx 1$. 

The other challenge is that realistic massive disks require a rather
high numerical resolution.  If one wants to resolve central features
below 100 pc in disks extending to at least 30 kpc, which must be able
to exchange angular momentum up to at least 100 kpc, one must have a
time or a scale range of about $10^4$.  Models with such a resolution
(with $N=2^{22} \approx 4.2\times 10^6$) are run on the GRAVITOR
Beowulf cluster at Geneva Observatory, with both particle-mesh code
and a parallelized version of the Barnes-Hut treecode.

\begin{figure}
\plottwo{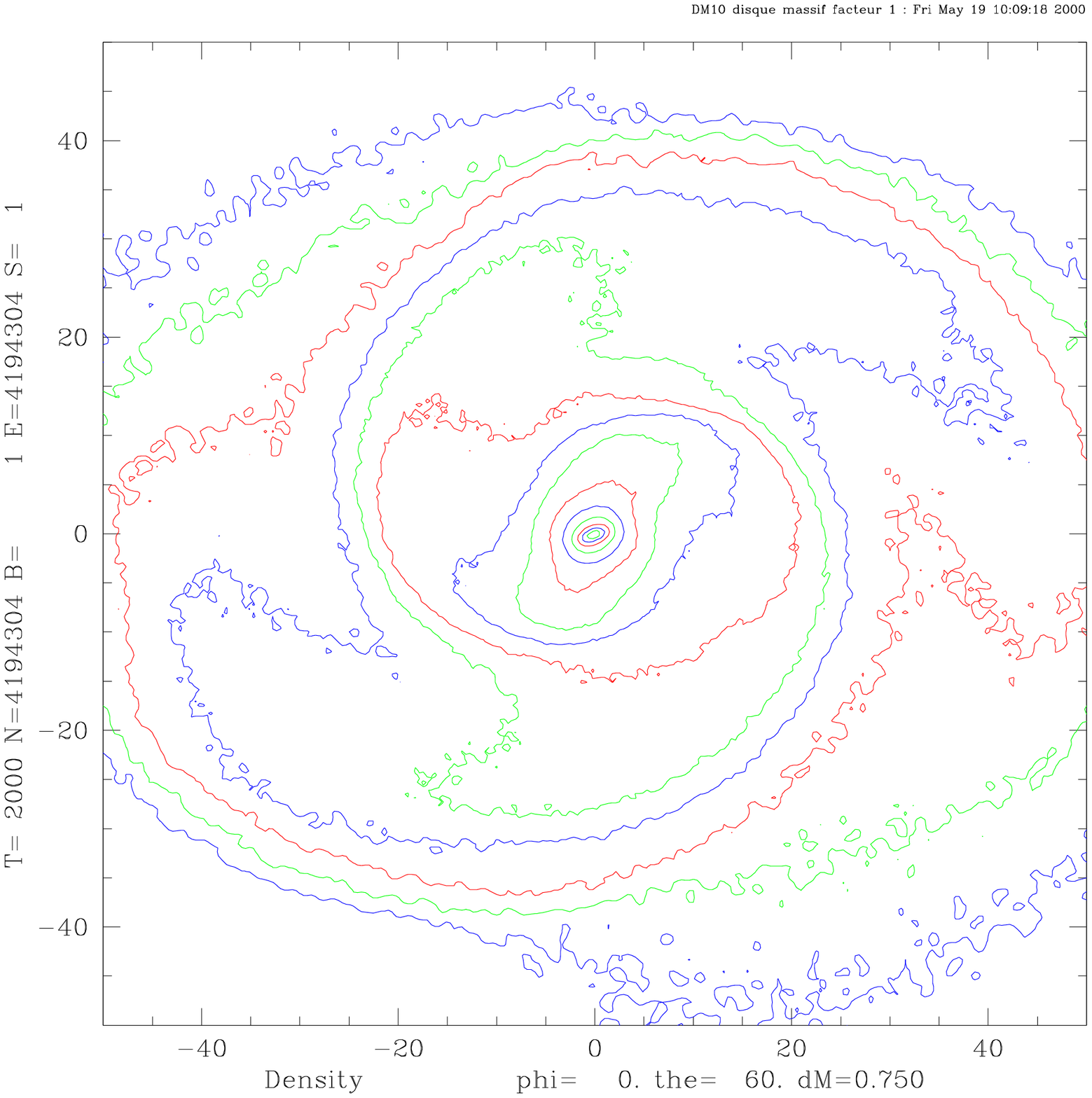}{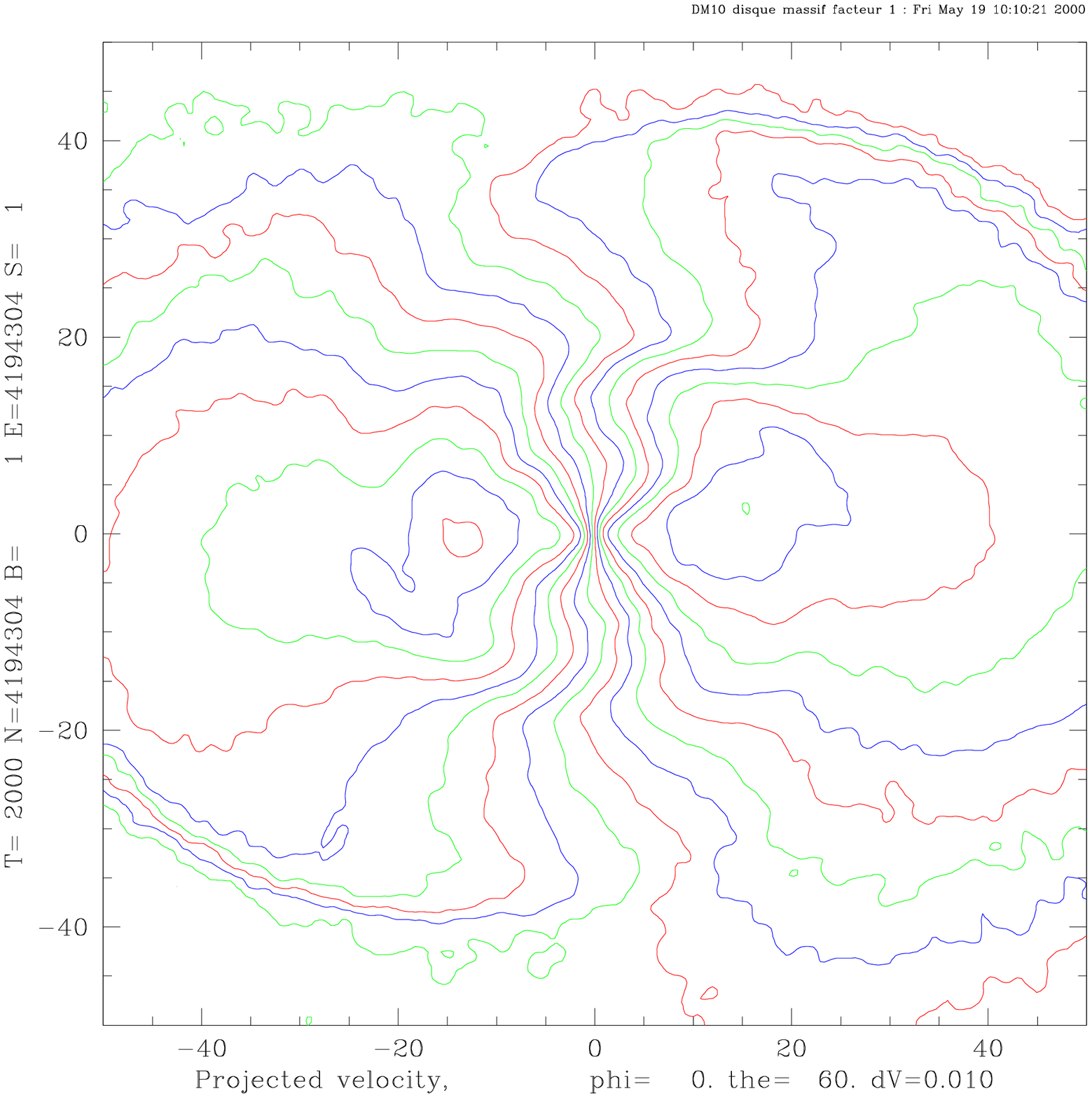}
\caption{A massive disk collisionless model inclined by $60\deg$ from
edge-on 2 Gyr after starting as a $Q\approx1$ axisymmetric
disk:\hfill\break 
Left: The total density contours at 0.75 mag interval.\hfill\break 
Right: The radial velocity contours at 10 km/s interval.}
\end{figure}

The initial conditions consist of a bulge, an optical exponential disk
and a heavy flaring disk proportional to the HI as in the Milky Way,
producing an almost flat rotation curve. Different initial velocity
dispersions of the components are explored, since the component 
temperatures are crucial for the overall stability.  Also different
``equations of state'' for treating the dissipation properties in the
cold gas are considered, in particular the degree of collisionality is
taken as a free parameter, from collisionless to viscous. 

Fig.~1 shows the morphology of the total mass distribution and the
corresponding velocity field of such a disk after 2 Gyr evolution.  No
energy dissipation is introduced in this example. A persistent double
bar can be seen near the center and prominent spiral arms in the
outer, dark matter dominated parts.

Overall, although rather flat, the models do not lead to clear
contradictions with observations.  Flaring maximum disks are not more
unstable that disks embedded in spheroidal halos, on the contrary,
dynamical friction between the visible and dark components vanishes.
The bars (single or double) are long lived, as well as the transient
but regenerated spiral arms, with characteristics times larger that a
few Gyr.  At 30 kpc radius the rotation period is of the order of 1
Gyr, therefore 10 Gyr time is short to obtain well virialized disks
there.  Actually, the outer disk radial asymmetries are
morphologically similar to the ones frequently observed in the outer
HI disks.

\section{Conclusions}
In the context of the current constraints about dark matter in
spirals, rotation supported dark halos are attractive because they do
allow simultaneously maximum baryonic inner disks and outer rotation
supported dark matter associated with HI.  The massive disk $N$-body
models run up to now, although rather extreme, do not show obvious
contradictions with observations.

\acknowledgements This work has been supported by the Swiss National Science 
Foundation.

\section*{Question:} 

\noindent
{\bf John Hibbard:} Do you think that your massive disks can explain
well-sepa\-ra\-ted mergers (nuclei separated by $>$ 5 kpc) with very long
($>100$ kpc) tails (e.g., NGC 4038/9, Arp 299)? If the tails contain an
order of magnitude more mass, they will carry away an order of
magnitude more angular momentum.

\noindent
{\bf Daniel Pfenniger:} Yes, massive rotating disks offer favorable
perspectives.  In fact, in the conventional picture of a baryonic disk
inside a hot dark halo, interacting galaxies should separate the low
and high angular momentum matter, and tails should be mostly baryonic.
But if dark and visible matter rotate the same, only the specific
angular momentum matters, and the exact mass contained in the tails
does not.  The tail mass density may be critical for the formation or
not of tidal dwarfs, and the dark matter content of the tidal dwarfs
provides a test about their initial composition.  Yet, if dark matter
is cold gas, one can expect a generous star formation in the tidal
dwarfs and a subsequent increase of the visible to dark matter ratio,
complicating the test.

\end{document}